\RequirePackage[l2tabu, orthodox]{nag}
\RequirePackage{snapshot}

\documentclass[9pt,onecolumn]{extarticle}

\makeatletter
\if@twocolumn
  \usepackage[dvips,a4paper,top=0.5in, bottom=0.5in, left=0.75in, right=0.5in,includefoot,heightrounded]{geometry}
\else
  \usepackage[dvips,a4paper,margin=1in,includefoot,heightrounded]{geometry}
\fi

\usepackage{srcltx}

\usepackage[russian,portuges,english]{babel}

\iflanguage{portuges}
    {\newcommand{\keywordname}{Palavras-chaves}}
    {\newcommand{\keywordname}{Keywords}}

\usepackage{amsmath}
\usepackage{amssymb}
\usepackage{amsfonts}
\usepackage{mathtools}
\usepackage{mathabx}

\usepackage{abstract}

\usepackage{graphicx}
\usepackage[usenames,dvipsnames,svgnames,x11names]{xcolor}
\usepackage{soul}
\usepackage{subfigure}

\usepackage{booktabs}

\usepackage{setspace}
\usepackage{flushend}

\usepackage{multicol}

\usepackage{cite}

\usepackage{hyperref}
\usepackage{breakurl}

\usepackage[normalem]{ulem}

\usepackage{enumerate}

\usepackage[noend]{algpseudocode}

\renewcommand{\And}{\mathbf{\ and\ }}
\newcommand{\Upto}{\mathbf{\ to\ }}

\newcommand{\False}{\textbf{false}}
\newcommand{\True}{\textbf{true}}
\newcommand{\Continue}{\textbf{continue}}

\newcommand*\BitAnd{\mathbin{\&}}
\newcommand*\BitOr{\mathbin{|}}
\newcommand*\BitXor{\mathbin{\oplus}}

\newcommand*\BitNot{\ensuremath{\mathord{\sim}}}

\newcommand*\OpAnd{\mathbin{\cdot}}
\newcommand*\OpOr{\mathbin{+}}

\newcommand*\OpNot{\bar}

\usepackage{algorithm}

\usepackage{listings}

\lstdefinelanguage{Julia}%
  {morekeywords={abstract,break,case,catch,const,continue,do,else,elseif,%
      end,export,false,for,function,immutable,import,importall,if,in,%
      macro,module,otherwise,quote,return,switch,true,try,type,typealias,%
      using,while},%
   sensitive=true,%
   morecomment=[l]\#,%
   morecomment=[n]{\#=}{=\#},%
   morestring=[s]{"}{"},%
   morestring=[m]{'}{'},%
}[keywords,comments,strings]%

\usepackage[short,12hr]{datetime}

\usepackage{siunitx}
\usepackage{tikz}
\usetikzlibrary{automata, positioning, arrows.meta}
\tikzset{
  draw=cyan!70!black,
  ->,                   %
  >=stealth,           %
  node distance=3cm,    %
}

       \usepackage{fouriernc}

\newcommand{\printtitle}{%
\makeatletter
\if@twocolumn

\twocolumn[%
  \maketitle
  \begin{onecolabstract}
    \myabstract
  \end{onecolabstract}
  \begin{center}
    \small
    \textbf{\keywordname}
    \\\medskip
    \mykeywords
  \end{center}
  \bigskip
]
\saythanks
\else
  \maketitle
  \begin{onecolabstract}
    \myabstract
  \end{onecolabstract}
  \begin{center}
    \small
    \textbf{\keywordname}
    \\\medskip
    \mykeywords
  \end{center}
  \bigskip
  \onehalfspacing
\fi
\makeatother
}

\author{%
R.~J.~Cintra%
\thanks{%
R. J. Cintra is with the
Signal Processing Group,
Department of Technology,
UFPE, Caruaru, Brazil.
E-mail: \url{rjdsc@de.ufpe.br}}
}

\title{%
A Note on the Conversion of Nonnegative Integers to the Canonical Signed-digit Representation}

\newcommand{\myabstract}{%
This note addresses the signed-digit representation of nonnegative binary integers.
Popular literature methods for
the conversion into the canonical signed-digit representation are reviewed and revisited.
A method based on string substitution is discussed.
}

\newcommand{\mykeywords}{%
Canonical signed-bit representation,
non-adjacent form,
binary numbers,
minimum-adder representation
}

\date{}

\begin{document}

\printtitle

\section{Introduction}

Let $x$ be an nonnegative integer.
The conventional binary representation of $x$
is
the $n$-uple
$
\mathbf{x}
=
\begin{bmatrix}
x_{n-1} & x_{n-2} & \cdots & x_i & \cdots & x_1 & x_0
\end{bmatrix}
$
such that
\begin{align}
x
=
\sum_{i=0}^{n-1}
2^i
\cdot
x_i
,
\end{align}
where
$n$ is the wordlength;
$0\leq i \leq n-1$;
$x_i \in \{0, 1\}$;
and
$x_i=0$, if $i<0$ or $i>n-1$,
For instance,
considering $n=4$,
the integer $x=7$
is mapped
into
$\mathbf{x} = \begin{bmatrix} 0&1&1&1 \end{bmatrix}.$
A comprehensive discussion on
conventional number systems
with non-binary base
is found in~\cite[p.~8]{hwang1979computer}.

The binary signed-digit number representation
of $\mathbf{x}$
is
an $(n+1)$-uple
$
\mathbf{y}
=
\begin{bmatrix}
y_n & y_{n-1} & y_{n-2} & \cdots & y_i & \cdots & y_1 & y_0
\end{bmatrix}
$
where
$y_i \in \{-1, 0, 1\}$
and
\begin{align}
x
=
\sum_{i=0}^n
2^i
\cdot
y_i
.
\end{align}
For example,
the number $x=7=8-4+2+1$
can be represented
as
\begin{align}
\mathbf{y}
=
\begin{bmatrix*}[r] 0 & 1 & -1 & 1 & 1  \end{bmatrix*}
.
\end{align}

Hereafter,
we might represent
$\mathbf{y}$
by two binary $(n+1)$-uples,
$
\mathbf{y^+}
=
\begin{bmatrix}
y^+_n & y^+_{n-1} & y^+_{n-2} & \cdots & y^+_i & \cdots & y^+_1 & y^+_0
\end{bmatrix}
$
and
$
\mathbf{y^-}
=
\begin{bmatrix}
y^-_n & y^-_{n-1} & y^-_{n-2} & \cdots & y^-_i & \cdots & y^-_1 & y^-_0
\end{bmatrix}
$,
indicating
the
positive
and
negative
coefficients of $\mathbf{y}$,
respectively.
Thus,
we have:
\begin{align}
y^+_i
&
=
\begin{cases}
1,      & \text{if $y_i=1$,}
\\
0,      & \text{otherwise,}
\end{cases}
\\
y^-_i
&
=
\begin{cases}
1,      & \text{if $y_i=-1$,}
\\
0,      & \text{otherwise,}
\end{cases}
\\
\mathbf{y}
&
=
\mathbf{y^+}
-
\mathbf{y^-}
.
\end{align}
Notice that
$\mathbf{y^+}$
and
$\mathbf{y^-}$
are conventional binary numbers;
not signed-digit numbers.

The conversion from a given binary number $\mathbf{x}$
to
its signed-digit representation
is not unique.
For example,
the following signed-digit representations
correspond to the same number ($x=7$):
\begin{align}
\begin{bmatrix*}[r]0 & 1 & -1 & 1 & 1  \end{bmatrix*}
,
\\
\begin{bmatrix*}[r]0 & 1 & 0 & -1 & 1  \end{bmatrix*}
,
\\
\begin{bmatrix*}[r]0 & 1 & 0 & 0 & -1  \end{bmatrix*}
.
\end{align}

Because the signed-digit representation is not unique,
representations that lead to minimal Hamming weight
might be advantageous.
For instance,
in~\cite[p.~12]{hwang1979computer},
a signed-digit representation is presented
which does not necessarily lead
to minimal weight.
The problem of minimal representation was addressed
in~\cite[p.~244--260]{reitwiesner1960binary}.
Reitwiesner showed that such a representation
not only exists but it is also unique~\cite[Sec.~8.3]{reitwiesner1960binary}.
Such minimal representation
is referred to as
the
canonical signed-digit representation.
A method to convert a generic signed-digit representation
to
the canonical signed-digit representation
is given in~\cite{tanaka2016efficient}.
However,
in this note,
we only address
methods
that convert numbers
from the
conventional binary representation
to
the canonical signed-digit representation.

We adopt the following symbols:
$\widebar{\phantom{x}}$,
$\OpAnd$,
$\BitNot$,
$\BitAnd$,
$\BitOr$,
$\BitXor$,
$-$,
$\ast$,
and
$\div$
for
logical negation,
logical conjunction (and),
bitwise logical negation,
bitwise logical conjunction (and),
bitwise logical disjunction (or),
exclusive or (xor) exclusive or bitwise xor,
arithmetic subtraction,
arithmetic multiplication,
and
integer division
operations,
respectively.
The symbol $+$
denotes
both
logical disjunction (or)
and
arithmetic addition,
being the meaning clear from the context.

This note is structured as follows.
Section~\ref{section-literature}
reviews popular methods in the literature.
In Section~\ref{section-revisited},
we revisit some of the discussed literature methods,
offering
comments and remarks.
Section~\ref{section-string}
describes
an encoding method based on
string substitution.
Section~\ref{section-performance}
discusses the performance results.
The note is concluded in Section~\ref{section-conclusion}.

\section{Literature Review}
\label{section-literature}

In this section,
we review
some methods for canonical signed-digit conversion.

\subsection{Reitwiesner's Method}

In~\cite[p.~252]{reitwiesner1960binary},
Reitwiesner proposed
a conversion from binary to the minimal signed-digit representation;
the method
is given in Algorithm~\ref{algo-reitwiesner1960binary-p252}.
The main computational bottleneck of this
algorithm
resides in the evaluation of recurrence
that generates the sequence
$g_i$, $i=0,1,\ldots,n$.

\begin{algorithm}[t]
\caption{Reitwiesner canonical signed-digit conversion algorithm in~\cite[p.~252]{reitwiesner1960binary}}
\label{algo-reitwiesner1960binary-p252}

\begin{algorithmic}[1]

\Procedure{reitwiesner}{$\mathbf{x}$}

\State $g_{-1} \gets 0$
\State $i \gets 0$

\While{$i \leq n$}

  \State $t_i \gets x_i \BitXor x_{i-1}$

  \State $g_i \gets \OpNot g_{i-1} \OpAnd t_i$

  \State $y_i \gets (1-2 \ast x_{i+1}) \ast g_i$
  \Comment{Arithmetical operation}

  \State $i \gets i+1$

\EndWhile

\State \Return $\mathbf{y}$

\EndProcedure

\end{algorithmic}

\end{algorithm}

The arithmetic expression
$1-2 \ast x_{i+1}$
is present in Reitwiesner's work~\cite[Eq.~8.6.9$\xi$, p.~252]{reitwiesner1960binary}
and
it is also adopted
in~\cite[p.~393]{avizienis1961signed}.
It
might be efficiently
computed by means of the following
identity:
$1-2 \ast x_{i+1} = \OpNot x_{i+1} - x_{i+1}$.

\subsection{Garner Method}

In~\cite[p.~164]{garner1966number},
Garner
showed a table with the rules for the canonical signed-digit code.
Such table and rules
are reproduced
in~\cite{peled1976hardware,hwang1979computer,ruiz2011efficient};
and we reproduce it yet again in Table~\ref{table-garner-rules}.
The coefficients $c_i$, $i=0,1,\ldots,n$,
represent the carry-out sequence
effected
by
the arithmetic addition
of
$\mathbf{x}$
and
$\mathbf{x}\div 2$.
Appendix~\ref{app-carryout}
offers a brief review on the carry-out computation.
Algorithm~\ref{algo-getcarry}
shows how to obtain such sequence
from the output of the full-adder operation.

From Garner's work~\cite{garner1966number},
it is not clear whether
the discussed coding rules
are originally
due to him.
Reitwiesner's work~\cite{reitwiesner1960binary}
presents three tables relating
the bit values of
$x_{i+i}$ and $x_i$
to several other auxiliary sequences
and
ultimately
to the encoded
signed-digit coefficients.
However,
no clear-cut table
as shown in~\cite[p.~164]{garner1966number}
was found
in~\cite{reitwiesner1960binary}.
Despite
the fact that
an explicit
algorithm for the rules in Table~\ref{table-garner-rules}
was not given in~\cite{garner1966number}---in the absence of better sources---we
refer to the algorithms immediately
derived from these rules
as the Garner algorithms.

\begin{table}[t]
\centering
\caption{Rules for converting conventional binary numbers
to canonical signed-digit representation}
\label{table-garner-rules}

\begin{tabular}{cccccc}
\toprule
$c_i$ & $x_{i+1}$ & $x_i$ & $c_{i+1}$ & $y^+_i$ & $y^-_i$
\\
\midrule
0 & 0 & 0 & 0 & 0 & 0\\
0 & 1 & 0 & 0 & 0 & 0\\
0 & 0 & 1 & 0 & 1 & 0\\
0 & 1 & 1 & 1 & 0 & 1\\
1 & 0 & 0 & 0 & 1 & 0\\
1 & 1 & 0 & 1 & 0 & 1\\
1 & 0 & 1 & 1 & 0 & 0\\
1 & 1 & 1 & 1 & 0 & 0\\
\bottomrule

\end{tabular}

\end{table}

The conversion rules (Table~\ref{table-garner-rules})
were algorithmically
described in at least two ways.
One approach
expresses the rules
in terms of arithmetic operations
as shown in~\cite[Eq.~5--6]{avizienis1961signed}
and~\cite[p.~150]{hwang1979computer}.
This method is shown in Algorithm~\ref{algo-garner1966number-arithmetic}.

\begin{algorithm}[t]
\caption{Garner canonical signed-digit conversion algorithm
as described in~\cite[Eq.~5--6]{avizienis1961signed}
and~\cite[p.~150]{hwang1979computer}}
\label{algo-garner1966number-arithmetic}

\begin{algorithmic}[1]

\Procedure{garnerArith}{$\mathbf{x}$}

\State $\mathbf{c} \gets \Call{getCarry}{\mathbf{x}, \mathbf{x}\div 2}$

\State $i \gets 0$

\While{$i \leq n$}
  \State $y_i \gets x_i + c_i - 2 \ast c_{i+1}$
  \Comment{Arithmetic operations}

    \State $i \gets i+1$
\EndWhile

\State \Return $\mathbf{y}$

\EndProcedure

\end{algorithmic}

\end{algorithm}

Another approach
consists
of
inspecting the truth table in~Table~\ref{table-garner-rules}
to
obtain
the logic expression
for
$y^+_i$ and $y^-_i$
in terms of
$c_i$, $x_{i+1}$, and $x_i$.
By doing so,
we have that
\begin{align}
\label{equation-garner-logic-first}
y^+_i
&=
\OpNot c_i \OpAnd \OpNot x_{i+1} \OpAnd x_i
\OpOr
c_i \OpAnd \OpNot x_{i+1} \OpNot \OpAnd x_i
\\
&=
\OpNot x_{i+1}
\OpAnd
(
c_i \BitXor x_i
)
,
\\
y^-_i
&=
\OpNot c_i \OpAnd x_{i+1} \OpAnd x_i
\OpOr
c_i \OpAnd x_{i+1} \OpAnd \OpNot x_i
\\
\label{equation-garner-logic-last}
&=
x_{i+1}
\OpAnd
(
c_i \BitXor x_i
)
.
\end{align}
Such logic-based formalism
coincides
with
the description
shown
in~\cite[Eq.~8--10]{ruiz2011efficient},
where~\cite{peled1976hardware,hwang1979computer}
are given as references.
However,
Garner's work~\cite{garner1966number}
antedates both~\cite{peled1976hardware} and~\cite{hwang1979computer}.
Peled~\cite{peled1976hardware}
points to~\cite{reitwiesner1960binary,garner1966number} as
primary sources;
whereas
Hwang~\cite{hwang1979computer} does not provide a reference.

\begin{algorithm}
\caption{Rule-based canonical signed-digit conversion algorithm}
\label{algo-garner}

\begin{algorithmic}[1]

\Procedure{garnerLogic}{$\mathbf{x}$}

\State $\mathbf{c} \gets \Call{getCarry}{\mathbf{x}, \mathbf{x}\div 2}$

\State $i \gets 0$

\While{$i \leq n$}
  \State $d_i \gets c_i \BitXor x_i$
  \State $y^+_i \gets \OpNot x_{i+1} \OpAnd d_i $
  \State $y^-_i \gets         x_{i+1} \OpAnd d_i $
  \State $y_i \gets y^+_i - y^-_i$
  \State $i \gets i+1$
\EndWhile

\State \Return $\mathbf{y}$

\EndProcedure

\end{algorithmic}

\end{algorithm}

\subsection{NAF Algorithm}

Referring to
the canonical signed-digit representation
as non-adjacent form (NAF),
Hankerson~\emph{et~al.}
present
an algorithm
for computing
the NAF of a positive integer in~\cite[p.~98]{hankerson2004guide}.
Algorithm~\ref{algo-hankerson2004guide-p98}
shows the associated pseudocode.

\begin{algorithm}
\caption{NAF algorithm}
\label{algo-hankerson2004guide-p98}

\begin{algorithmic}[1]

\Procedure{naf}{$x$}

\State $i \gets 0$

\While{$x \geq 1$}

  \If{$x$ is odd}
    \State $y_i \gets 2 - x \pmod{4}$
    \State $x \gets x - y_i$
  \Else
    \State $y_i \gets 0$
  \EndIf

  \State $x \gets x \div 2$
  \State $i \gets i + 1$

\EndWhile

\State \Return $\mathbf{y}$

\EndProcedure

\end{algorithmic}

\end{algorithm}

\subsection{bin2naf Algorithm}

In~\cite[p.~61]{arndt2011matters},
the canonical signed-digit
representation
is referred to as
sparse signed binary representation
or---as in~\cite{hankerson2004guide}---as NAF.
An algorithm to convert
the conventional binary representation
to the nonadjacent form
is also furnished,
which is shown in Algorithm~\ref{algo-arndt2011matters-p62}.

\begin{algorithm}
\caption{Binary to NAF algorithm}
\label{algo-arndt2011matters-p62}

\begin{algorithmic}[1]

\Procedure{bin2naf}{$\mathbf{x}$}

\State $\mathbf{h} \gets \mathbf{x}\div2$
\State $\mathbf{t} \gets \mathbf{x} + \mathbf{h}$

\State $\mathbf{d} \gets \mathbf{h} \BitXor \mathbf{t}$

\State $\mathbf{y}^+ \gets \mathbf{t} \BitAnd \mathbf{d}$
\State $\mathbf{y}^- \gets \mathbf{h} \BitAnd \mathbf{d}$

\State $\mathbf{y} \gets \mathbf{y}^+ - \mathbf{y}^-$

\State \Return $\mathbf{y}$

\EndProcedure

\end{algorithmic}

\end{algorithm}

\subsection{Ruiz-Granda Algorithm}

Ruiz and Granda
introduced in~\cite{ruiz2011efficient}
a method for canonical signed-digit coding.
The method is based on
two auxiliary recursive sequences.
Algorithms~\ref{algo-ruiz2011efficient-hk} and~\ref{algo-ruiz2011efficient}
describe the procedure.

\begin{algorithm}[t]
\caption{Ruiz-Granda $\mathbf{h}$ and $\mathbf{k}$ sequences}
\label{algo-ruiz2011efficient-hk}

\begin{algorithmic}[1]

\Procedure{getHK}{$\mathbf{x}$}

\State $h_{-1} \gets 0$
\State $k_{-1} \gets 0$
\State $i \gets 0$

\While{$i \leq n$}
  \If{$i$ is even}
    \State $h_i \gets x_i \OpOr h_{i-1}$
    \State $k_i \gets x_i \OpAnd k_{i-1}$
  \Else
    \State $h_i \gets x_i \OpAnd h_{i-1}$
    \State $k_i \gets x_i \OpOr k_{i-1}$
  \EndIf

  \State $i \gets i + 1$
\EndWhile

\State \Return $\mathbf{h}$, $\mathbf{k}$

\EndProcedure

\end{algorithmic}

\end{algorithm}

\begin{algorithm}[t]
\caption{Ruiz-Granda Algorithm}
\label{algo-ruiz2011efficient}

\begin{algorithmic}[1]

\Procedure{ruizGranda}{$\mathbf{x}$}

\State $\mathbf{h},\mathbf{k} \gets \Call{getHK}{\mathbf{x}}$

\State $i \gets 0$

\While{$i \leq n$}

  \If{$i$ is even}
    \State $d_i \gets h_i \OpAnd \OpNot k_i$
  \Else
    \State $d_i \gets \OpNot h_i \OpAnd k_i$
  \EndIf

  \State $y^+_i \gets \OpNot x_{i+1} \OpAnd d_i$
  \State $y^-_i \gets x_{i+1} \OpAnd d_i$
  \State $y_i \gets y^+_i - y^-_i$

  \State $i \gets i+1$

\EndWhile

\State \Return $\mathbf{y}$

\EndProcedure

\end{algorithmic}

\end{algorithm}

\section{Revisited Methods}
\label{section-revisited}

In this section,
we revisit some literature methods,
aiming at
deriving remarks and relations
among them.

\subsection{Modified Reitwiesner's Method}

The recursion required by Reitwiesner's method
can be obtained
by means of the carry-out recursion.
It can be shown that the sequence
$g_i$, $i=0,1,\ldots,n$,
(Algorithm~\ref{algo-reitwiesner1960binary-p252})
satisfies
the following relation:
\begin{align}
\label{eq-g-seq-efficient}
g_i
=
c_{i+1}
\BitXor
x_i
,
\end{align}
where
$c_i$, $i=0,1,\ldots,n$,
is the particular carry-out sequence
generated by
the addition
$\mathbf{x} + 2 \ast \mathbf{x}$.
By applying the above considerations
to
Algorithm~\ref{algo-reitwiesner1960binary-p252},
we obtain
Algorithm~\ref{algo-reitwiesner-modified}.

\begin{algorithm}[t]
\caption{Modified Reitwiesner canonical signed-digit conversion algorithm}
\label{algo-reitwiesner-modified}

\begin{algorithmic}[1]

\Procedure{reitwiesnerModified}{$\mathbf{x}$}

\State $\mathbf{c} \gets \Call{getCarry}{\mathbf{x},2 \ast \mathbf{x}}$

\State $i \gets 0$

\While{$i \leq n$}

  \State $g_i \gets c_{i+1} \BitXor x_i$

  \State $y^+_i \gets \OpNot x_{i+1} \OpAnd g_i$
  \State $y^-_i \gets x_{i+1} \OpAnd g_i$
  \State $y_i \gets y^+_i - y^-_i$

  \State $i \gets i+1$

\EndWhile

\State \Return $\mathbf{y}$

\EndProcedure

\end{algorithmic}

\end{algorithm}

\subsection{bin2naf Revisited}

Algorithm~\ref{algo-arndt2011matters-p62}
is very similar to
Algorithm~\ref{algo-garner}.
To show this,
let us recast
Algorithm~\ref{algo-garner}
in
terms of tuple operations,
instead of individual coefficient operations.
The result is
Algorithm~\ref{algo-garner1966number-2}.

\begin{algorithm}
\caption{Reformatted Algorithm~\ref{algo-garner}}
\label{algo-garner1966number-2}

\begin{algorithmic}[1]

\Procedure{garner}{$\mathbf{x}$}

\State $\mathbf{h} \gets \mathbf{x}\div2$
\State $\mathbf{c} \gets \Call{getCarry}{\mathbf{x}, \mathbf{h}}$

\State $\mathbf{d} \gets \mathbf{c} \BitXor \mathbf{x}$

\State $\mathbf{y}^+ \gets \BitNot \mathbf{h} \BitAnd \mathbf{d}$
\State $\mathbf{y}^- \gets         \mathbf{h} \BitAnd \mathbf{d}$

\State $\mathbf{y} \gets \mathbf{y}^+ - \mathbf{y}^-$

\State \Return $\mathbf{y}$

\EndProcedure

\end{algorithmic}

\end{algorithm}

First,
notice that the tuple $\mathbf{d}$
in Algorithm~\ref{algo-arndt2011matters-p62}
is
equal to the tuple $\mathbf{d}$
from Algorithm~\ref{algo-garner1966number-2},
as shown below:
\begin{align}
\mathbf{d}
&=
\mathbf{h}\BitXor\mathbf{t}
\label{equation-d=h-xor-t}
\\
&=
\mathbf{h}\BitXor
(\mathbf{c}\BitXor\mathbf{x}\BitXor\mathbf{h})
\\
&=
\mathbf{h}\BitXor
\mathbf{h}\BitXor\mathbf{c}\BitXor\mathbf{x}
\\
&=
\mathbf{c}\BitXor\mathbf{x}
\\
&=
\mathbf{d}
\label{eq-d=x-xor-c}
.
\end{align}
Now let us shown that both $\mathbf{y}^+$ are also identical.
Notice that
$\mathbf{t} =
\mathbf{c} \BitXor \mathbf{x} \BitXor \mathbf{h}
=
\mathbf{d} \BitXor \mathbf{h}$
(see \eqref{eq-d=x-xor-c}).
From Algorithm~\ref{algo-arndt2011matters-p62},
we have that:
\begin{align}
\mathbf{y}^+
&=
\mathbf{t} \BitAnd \mathbf{d}
\\
&=
(\mathbf{d} \BitXor \mathbf{h}) \BitAnd \mathbf{d}
\\
&=
\Big(
(\mathbf{d}\BitAnd\BitNot\mathbf{h})
\BitOr
(\BitNot\mathbf{d}\BitAnd\mathbf{h})
\Big)
\BitAnd \mathbf{d}
\\
&=
\BitNot\mathbf{h}\BitAnd\mathbf{d}
\\
&=
\mathbf{y}^+
,
\end{align}
which is the expression for
$\mathbf{y}^+$ in Algorithm~\ref{algo-garner1966number-2}.
As for $\mathbf{y}^-$,
the logical expressions are already plainly identical
in both algorithms.

\subsection{Garner Algorithm Revisited}

Algorithm~\ref{algo-garner1966number-2}
can be further simplified.
First,
from~\eqref{equation-d=h-xor-t},
recall that
$
\mathbf{d}
=
\mathbf{h} \BitXor \mathbf{t}
$.
From Algorithm~\ref{algo-arndt2011matters-p62},
we have that:
\begin{align}
\mathbf{y}^+
&=
\mathbf{t} \BitAnd \mathbf{d}
\\
&=
\mathbf{t} \BitAnd ( \mathbf{h} \BitXor \mathbf{t} )
\\
&=
\BitNot \mathbf{h} \BitAnd \mathbf{t}
.
\end{align}
Similarly,
from
Algorithm~\ref{algo-garner1966number-2},
the following is true:
\begin{align}
\mathbf{y}^+
&=
\BitNot \mathbf{h} \BitAnd \mathbf{d}
\\
&=
\BitNot \mathbf{h} \BitAnd ( \mathbf{h} \BitXor \mathbf{t} )
\\
&=
\BitNot \mathbf{h} \BitAnd \mathbf{t}
.
\end{align}
And finally,
from Algorithms~\ref{algo-arndt2011matters-p62}
and~\ref{algo-garner1966number-2},
we obtain:
\begin{align}
\mathbf{y}^-
&=
\mathbf{h} \BitAnd \mathbf{d}
\\
&=
\mathbf{h} \BitAnd ( \mathbf{h} \BitXor \mathbf{t} )
\\
&=
\mathbf{h} \BitAnd \BitNot \mathbf{t}
.
\end{align}

By applying the above results
back into Algorithm~\ref{algo-garner1966number-2},
we obtain
Algorithm~\ref{algo-garner1966number-3}.

\begin{algorithm}
\caption{Revisited Garner canonical signed-digit conversion algorithm}
\label{algo-garner1966number-3}

\begin{algorithmic}[1]

\Procedure{garnerRevisited}{$\mathbf{x}$}

\State $\mathbf{h} \gets \mathbf{x}\div2$
\State $\mathbf{t} \gets \mathbf{x} + \mathbf{h}$

\State $\mathbf{y}^+ \gets \BitNot \mathbf{h} \BitAnd \mathbf{t}$
\State $\mathbf{y}^- \gets \mathbf{h} \BitAnd \BitNot \mathbf{t}$

\State $\mathbf{y} \gets \mathbf{y}^+ - \mathbf{y}^-$

\State \Return $\mathbf{y}$

\EndProcedure

\end{algorithmic}

\end{algorithm}

\section{String Substitution Method}
\label{section-string}

In this section,
we describe a direct method
based on string substitution.

\subsection{Pattern Finding}

Following \cite[p.~611, Answer~34]{knuth1997art}
and
\cite[Fig.~1]{ruiz2011efficient},
a moment's reflection
shows that
the canonical signed-digit representation
can be achieved
by direct pattern substitution.
Let us denote the pattern $00\mathbf{w}11$ as a `w-block',
where~$\mathbf{w}$ is
either empty
or
the longest sequence of zeros and ones
such that
(i) it has a leading~1 at the leftmost position
and
(ii) no consecutive zeros are present.
The canonical signed-digit
representation
consists
of the following substitution:
\begin{align}
\ldots\boxed{00\mathbf{w}11}\ldots
\to
\ldots\boxed{01[-\overline{\mathbf{w}}]0\bar{1}}\ldots
,
\end{align}
where
$\bar{1} \triangleq -1$;
$[-\overline{\mathbf{w}}]$
operates over the digits of $\mathbf{w}$
according to:
$0\to\bar{1}$
and
$1\to0$;
and
$[-\overline{\mathbf{w}}]=\emptyset$,
if
$\mathbf{w}=\emptyset$.
All other patterns that do not match
the `w-block' format
are
simply copied to the output
without any change.
For example,
\begin{align}
0000\boxed{00111}000  \to  & 0000\boxed{0100\bar{1}}000
,
\\
\boxed{001010111}000  \to  & \boxed{010\bar{1}0\bar{1}00\bar{1}}000
,
\\
0010\boxed{00111}001  \to  & 0010\boxed{0100\bar{1}}001
,
\\
0010\boxed{00111011}  \to  & 0010\boxed{01000\bar{1}0\bar{1}}
,
\\
0000\boxed{0011}0000  \to  & 0000\boxed{010\bar{1}}0000
.
\end{align}
The above last example
illustrates the case in which $\mathbf{w}$ is empty.
The sequence of numbers in the `w-block' form is~\cite[A380358]{oeis}:
\begin{center}
  3,   7,  11,  15,  23,  27,  31,  43,  47,  55,  59,  63,  87,  91,  95, 107,
111, 119, 123, 127, 171, 175, 183, 187, 191, 215, 219, 223, 235, 239, 247, 251,
255, 343, 347, 351, 363, 367, 375, 379, 383, 427, 431, 439, 443, 447, 471, 475,
479, 491, 495, 503, 507, 511, 683, 687, 695, 699, 703, 727, 731, 735, 747, 751,
759, 763, 767, 855, 859, 863, 875, 879, 887, 891, 895, 939, 943, 951, 955, 959,
983, 987, 991, 1003, 1007, 1015, 1019, 1023, \ldots
\end{center}
This sequence is related to sequences
A003754
and
A247648
listed in~\cite{oeis}
according
to
$4 \cdot \mathrm{A003754}(n) + 3$
and
$2 \cdot \mathrm{A247648}(n) + 1$.
A comparable sequence is shown in~\cite[p.~21]{smith1958shortcut}.

One way of mathematically
describing
the
above
pattern substitution
consists
of
considering
(i)~a
two-bit sliding window
moving from right to left
that scans
two consecutive digits
$x_{i+1}$ and $x_i$
and
(ii)~an flag bit $f_i$
($f_0 = 0$)
to indicate whether the
sliding window is over
a `w-block'
or not.
Two consecutive ones mark the start of the `w-block' pattern
($f_{i+1}=1$);
whereas
two consecutive zeros mark its end
($f_{i+1}=0$).
For each $i$,
the resulting
canonical signed-digit number
is
$y_i$.

Algorithm~\ref{algo-string-0}
directly realizes
the discussed string-based approach.
Algorithm~\ref{algo-string-1}
is
an alternative version of~Algorithm~\ref{algo-string-0}.

\begin{algorithm}[t]
\caption{String substitution canonical signed-digit conversion algorithm (32-bit version)}
\label{algo-string-0}

\begin{algorithmic}[1]

\Procedure{\detokenize{string_0}}{$\mathbf{x}$}

\State $f \gets \False$

\For{$i = 0 \Upto 32$}

  \State $\mathbf{p} \gets [x_{i+1},x_i]$

  \If{$f$ is \False}

    \If{$\mathbf{p} = [0,1]$}
      \State $y_i \gets 1$
    \ElsIf{$\mathbf{p} = [1,1]$}
      \State $y_i \gets -1$
      \State $f \gets \True$
    \Else
      \State $y_i \gets 0$
    \EndIf

  \Else

    \If{$\mathbf{p} = [0,0]$}
      \State $y_i \gets 1$
      \State $f \gets \False$
    \ElsIf{$\mathbf{p} = [1,0]$}
      \State $y_i \gets -1$
    \Else
      \State $y_i \gets 0$
    \EndIf

  \EndIf

\EndFor

\State \Return $\mathbf{y}$

\EndProcedure

\end{algorithmic}

\end{algorithm}

\begin{algorithm}
\caption{Alternative version of the string substitution canonical signed-digit conversion algorithm (32-bit version)}
\label{algo-string-1}

\begin{algorithmic}[1]
\Procedure{\detokenize{string_1}}{$\mathbf{x}$}

\State $\mathbf{y} \gets \mathbf{x}$

\State $i \gets 0$

\While{$i \leq 32$}

    \If{$y_i = 0$}
        \State $i \gets i + 1$
        \State \Continue
    \EndIf

    \State $j \gets 1$

    \While{$(y_{i+j}=1 \And y_{i+j-1}=1)$}
        \State $j \gets j + 1$
    \EndWhile

    \If{$j > 1$}
        \State $y_{i+j} \gets 1$

        \For{$k \gets i+1 \Upto i+j-1$}
          \State $y_k \gets 0$
        \EndFor

        \State $y_i \gets -1$
    \EndIf

    \State $i \gets i+j$

\EndWhile

\State \Return $\mathbf{y}$

\EndProcedure

\end{algorithmic}
\end{algorithm}

\subsection{Proof}

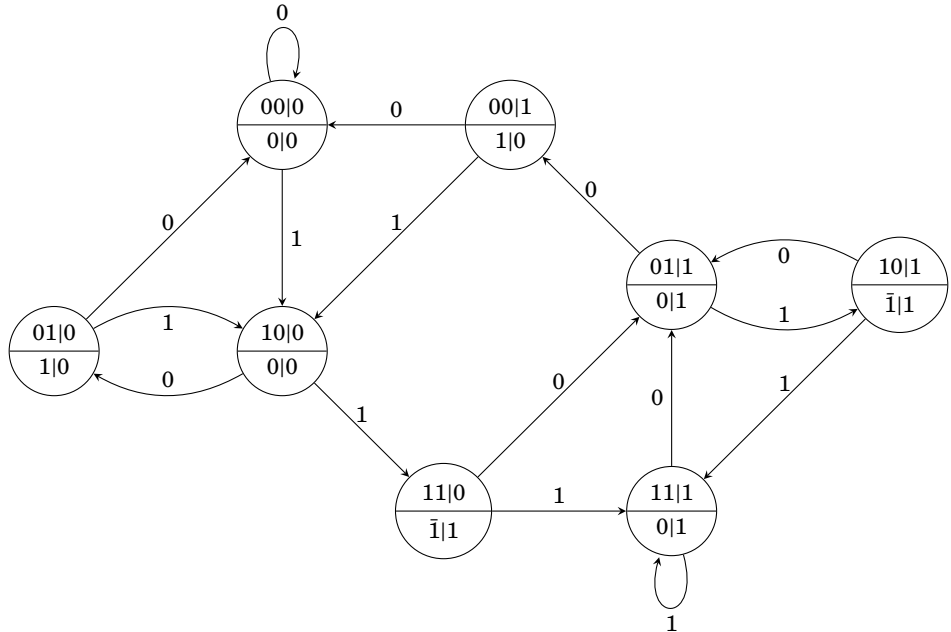
\begin{figure*}[t]
\centering
\begin{tikzpicture}
\node[state, circle split,]
  (z0)
  {$00|0$ \nodepart{lower} $0|0$};
\node[state, circle split, below of=z0]
  (z4)
  {$10|0$ \nodepart{lower} $0|0$};
\node[state, circle split, left of=z4]
  (z2)
  {$01|0$ \nodepart{lower} $1|0$};
\node[state, circle split, below right of=z4]
  (z6)
  {$11|0$ \nodepart{lower} $\bar{1}|1$};
\node[state, circle split, right of=z6]
  (z7)
  {$11|1$ \nodepart{lower} $0|1$};
\node[state, circle split, above of=z7]
  (z3)
  {$01|1$ \nodepart{lower} $0|1$};
\node[state, circle split, right of=z3]
  (z5)
  {$10|1$ \nodepart{lower} $\bar{1}|1$};
\node[state, circle split, above left of=z3]
  (z1)
  {$00|1$ \nodepart{lower} $1|0$};

\draw
(z0) edge[loop above]        node{0} (z0)
(z0) edge[right]             node{1} (z4)
(z4) edge[bend left, above]  node{0} (z2)
(z4) edge[above]             node{1} (z6)
(z2) edge[above]             node{0} (z0)
(z2) edge[bend left, below]  node{1} (z4)
(z6) edge[above]             node{0} (z3)
(z6) edge[above]             node{1} (z7)
(z1) edge[above]             node{0} (z0)
(z1) edge[above]             node{1} (z4)
(z5) edge[bend right, below] node{0} (z3)
(z5) edge[above]             node{1} (z7)
(z3) edge[above]             node{0} (z1)
(z3) edge[bend right, above] node{1} (z5)
(z7) edge[left]              node{0} (z3)
(z7) edge[loop below]        node{1} (z7)
;
\end{tikzpicture}
\caption{State diagram for
the conversion to the canonical signed-digit representation.
State $x_{i+1}x_i|f_i$ and output $y_i|f_{i+1}$.}
\label{fig-state-diagram}
\end{figure*}

\begin{table}[t]
\centering
\caption{Rules for converting conventional binary numbers
to canonical signed-digit representation}
\label{table-rules}

\begin{tabular}{ccccccc}
\toprule
$x_{i+1}$ & $x_i$ & $f_i$ & $y_i$ & $y^+_i$ & $y^-_i$ & $f_{i+1}$
\\
\midrule
0 & 0 & 0 & 0 & 0 & 0 & 0 \\
0 & 1 & 0 & 1 & 1 & 0 & 0 \\
1 & 0 & 0 & 0 & 0 & 0 & 0 \\
1 & 1 & 0 & $\bar{1}$ & 0 & 1 & 1 \\
\midrule
0 & 0 & 1 & 1 & 1 & 0 & 0 \\
0 & 1 & 1 & 0 & 0 & 0 & 1 \\
1 & 0 & 1 & $\bar{1}$ & 0 & 1 & 1 \\
1 & 1 & 1 & 0 & 0 & 0 & 1 \\
\bottomrule
\end{tabular}

\end{table}

The pattern substitution rules
imply the finite-state machine
shown in~Figure~\ref{fig-state-diagram},
which
translates
into
Table~\ref{table-rules}.
By inspecting this truth table (Table~\ref{table-rules}),
we can obtain
the logical expressions
for
$y^+_i$,
$y^-_i$,
and
$f_i$,
as follows:
\begin{align}
y_i^+
&=
\OpNot x_{i+1} \OpAnd x_i \OpAnd \OpNot f_i
+
\OpNot x_{i+1} \OpAnd \OpNot x_i \OpAnd f_i
\\
&=
\OpNot x_{i+1}
\OpAnd
(
x_i
\BitXor
f_i
)
,
\\
y_i^-
&=
x_{i+1} \OpAnd x_i \OpAnd \OpNot f_i
+
x_{i+1} \OpAnd \OpNot x_i \OpAnd f_i
\\
&=
x_{i+1}
\OpAnd
(
x_i
\BitXor
f_i
)
,
\\
f_{i+1}
&=
x_{i+1} \OpAnd x_i \OpAnd \OpNot f_i
\OpOr
\OpNot x_{i+1} \OpAnd x_i \OpAnd f_i
\OpOr
x_{i+1} \OpAnd x_i \OpAnd f_i
\OpOr
x_{i+1} \OpAnd \OpNot x_i \OpAnd f_i
\\
&=
x_{i+1} \OpAnd x_i
\OpOr
\OpNot x_{i+1} \OpAnd x_i \OpAnd f_i
\OpOr
x_{i+1} \OpAnd \OpNot x_i \OpAnd f_i
\\
&=
x_{i+1} \OpAnd x_i
\OpOr
f_i
\OpAnd
\Big(
\OpNot x_{i+1} \OpAnd x_i
\OpOr
x_{i+1} \OpAnd \OpNot x_i
\Big)
\\
&=
x_{i+1} \OpAnd x_i
\OpOr
f_i
\OpAnd
(x_{i+1} \BitXor x_i)
.
\end{align}
One can recognize
that
the above expressions
for
$y^+_i$
and
$y^-_i$
are
identical
to
the expressions
\eqref{equation-garner-logic-first}--\eqref{equation-garner-logic-last}.
Moreover,
the flag bit $f_i$
and
the carry-out sequence $c_i$
from a full adder
have the same logical expression.
Therefore,
the discussed
string substitution algorithm
is mathematically equivalent to
Algorithms~\ref{algo-arndt2011matters-p62},
\ref{algo-garner1966number-2},
and~\ref{algo-garner1966number-3}.

\section{Performance Evaluation and Discussion}
\label{section-performance}

\subsection{Coding}

The algorithms listed in the previous sections
were implemented in Julia language~\cite{juliaproject2024julia}.
As much as we could,
the sought codes were
based on
bitwise operations
for
higher performance.
The programs usually follow the direct specification
from their respective pseudocodes.
The resulting codes were checked against all 32-bit inputs.
The method \textsc{bin2naf} was adopted as the reference standard.

\subsection{Assessment}

Table~\ref{table-comparison}
displays the benchmark statistical summary
as evaluated by the
Julia BenchmarkTools~\cite{2025benchmarktoolsjl}.
The Mersenne Twister pseudorandom number generator
was employed to
generate
uniform random integers
in the interval $[0, 2^{32}-1]$.
The same random data
were submitted
as input to each discussed algorithm.
Results were estimated
according to \qty{1000} samples of \qty{1000} evaluations each.
Multithreading was not employed.
The machine used for the performance measurements
possessed a
64-bit processor (Intel Core i5-10400F CPU)
running at an average clock of \qty{4.1}{\GHz}.

\begin{table*}
\centering
\sisetup{table-format = 2.3, table-alignment-mode = format}
\caption{Performance comparison of the discussed methods:
Statistics of execution time
(measured in \unit{\ns}).
Algorithms sorted by mean measurements.
See text for implementation details.}
\label{table-comparison}
\begin{tabular}
{
l
S[table-number-alignment = center]
S[table-number-alignment = center]
S[table-number-alignment = center]
S[table-number-alignment = center]
S[table-number-alignment = center]
}
\toprule
{Algorithm} & {Mininum} & {Maximum} & {Median} & {Mean} & {Std dev}
\\
\midrule
\textsc{garnerRevisited}
        &   2.403   &   20.401   &   2.419   &   2.445   &  0.570
\\
\textsc{garner}~\cite{garner1966number}
        &   2.405   &   18.652   &   2.421   &   2.447   &  0.563
\\
\textsc{bin2naf}~\cite{arndt2011matters}
        &   2.404   &   25.848   &   2.423   &   2.463   &  0.734
\\
\midrule
\textsc{reitwiesnerModified}
        &   2.646   &   44.773   &   2.710   &   2.737   &  0.905
\\
\textsc{\detokenize{string_1}}
        &   11.612   &   41.374  &  11.778   &  12.063   &  1.783
\\
\textsc{reitwiesner}~\cite{reitwiesner1960binary}
        &   23.426   &   65.160  &  24.009   &  24.347   &  2.088
\\
\textsc{naf}~\cite{hankerson2004guide}
        &   25.585   &   86.374  &  25.701   &  26.399   &  2.860
\\
\textsc{\detokenize{string_0}}
        &   29.162   &   70.727  &  29.187   &  29.676   &  2.929
\\
\bottomrule
\end{tabular}
\end{table*}

\subsection{Discussion}

The Julia implementation for
algorithms
\textsc{garner}
(Algorithm~\ref{algo-garner1966number-2}),
\textsc{bin2naf}
(Algorithm~\ref{algo-arndt2011matters-p62}),
and
\textsc{garnerRevisited}
(Algorithm~\ref{algo-garner1966number-3})
showed
essentially
the same performance.
The particular order in which these algorithms
appear in Table~\ref{table-comparison}
should not be taken as definite.
The estimated performance measurements
presented small random variations
in the simulation,
if the pseudorandom generator seed value
changes.
The measurements, however,
keep roughly unchanged.
The absence of
loop and branching structures
combined
with
low-level bitwise operations
ensured their relatively higher performance.
As shown in Section~\ref{section-revisited},
these three methods
have very similar structures.

Algorithms~\textsc{reitwiesner}
(Algorithm~\ref{algo-reitwiesner1960binary-p252})
and
\mbox{\textsc{ruizGranda}}
(Algorithm~\ref{algo-ruiz2011efficient})
are based on the evaluation of
specific
recursive sequences.
The former requires
the sequence~$\mathbf{g}$;
whereas
the latter
needs sequences~$\mathbf{h}$ and~$\mathbf{k}$.
The sequence~$\mathbf{g}$
could be connected
to the carry-out sequence
by means of~\eqref{eq-g-seq-efficient}.
Therefore,
$\mathbf{g}$~could be efficiently computed.
This resulted in the
\mbox{\textsc{reitwiesnerModified}} algorithm
(Algorithm~\ref{algo-reitwiesner-modified})

On the other hand,
we could not find an efficient
way to solve the recursions
required by
the \mbox{\textsc{ruizGranda}} algorithm.
Thus we resorted to
the direct approach
based on loop structures
(Algorithm~\ref{algo-ruiz2011efficient-hk}).
Unfortunately,
this approach
effected bottlenecks and resulted in
comparatively increased execution times
(%
Min.:~\qty{46.255}{\ns},
Max.:~\qty{105.083}{\ns},
Median:~\qty{47.503}{\ns},
Mean:~\qty{47.901}{\ns},
and
Standard deviation:~\qty{4.525}{\ns}).
Such
measurements
should be taken with care.
The relatively higher execution times
are primarily due to
our
failure to derive
an efficient implementation
for the computation of the required
recursive sequences
rather
than
to any intrinsic inefficiency in the algorithm itself.
As a consequence,
these results
were not included
in Table~\ref{table-comparison},
lest it would
be an unfair comparison
or
a misleading information.

The string substitution method
\textsc{\detokenize{string_0}}
(Algorithm~\ref{algo-string-0})
is comparable to
the method proposed in~\cite[Algorithm~5.2]{imam2004symbolic},
which
relies
on
a pattern search algorithm
(\texttt{FindPattern()},~\cite[Algorithm~5.1]{imam2004symbolic}).
We could not clearly identify from~\cite{imam2004symbolic}
how their pattern search algorithm operates,
which prevented our reproducing their algorithm.
Additionally,
the substitution scanner provided in~\cite{imam2004symbolic}
requires
a scanning window
encompassing
$x_{i+1}$, $x_i$, $x_{i-1}$, and $x_{i-2}$;
whereas
the discussed
scanner
reads
$x_{i+1}$ and $x_i$ only.

The alternative algorithm~\textsc{\detokenize{string_1}}
(Algorithm~\ref{algo-string-1})
could
perform
better
than
algorithms~\textsc{reitwiesner}
(Algorithm~\ref{algo-reitwiesner1960binary-p252})
and
\textsc{naf}
(Algorithm~\ref{algo-hankerson2004guide-p98}).
However,
it could not outcompete
the top performing methods---yet
again a statement
on the efficiency of the
full-adder
operation
in current processors.

\section{Conclusions}
\label{section-conclusion}

To the best of our knowledge,
the Garner Method as shown
in the revised form \textsc{garnerRevisited}
(Algorithm~\ref{algo-garner1966number-3})
is not listed in the literature.
We clarified the equivalence
of
methods
\textsc{garnerRevisited}
(Algorithm~\ref{algo-garner1966number-3}),
\textsc{bin2naf}
(Algorithm~\ref{algo-arndt2011matters-p62}),
and
\textsc{garner}
(Algorithm~\ref{algo-garner1966number-2}).
We emphasized
the role of the carry-out sequence,
derived from a full adder,
as a means to improve the performance
of \mbox{\textsc{reitwiesner}} algorithm,
resulting
in
\mbox{\textsc{reitwiesnerModified}}
algorithm
(Algorithm~\ref{algo-reitwiesner-modified}).

As far as we know,
the string substitution
algorithms
as shown in
\mbox{\textsc{\detokenize{string_0}}}
(Algorithm~\ref{algo-string-0})
and
\textsc{\detokenize{string_1}}
(Algorithm~\ref{algo-string-1})
are not archived in the literature,
albeit comparable methods are available.

For future work,
we aim at providing
an efficient way for computing
the sequences $\mathbf{h}$ and $\mathbf{k}$
required by Algorithm~\ref{algo-ruiz2011efficient}.
In conclusion,
due to its simplicity and high performance,
the revisited Garner algorithm
(\textsc{garnerRevisited},
Algorithm~\ref{algo-garner1966number-3})
appears as the recommended algorithm.

{\small
\singlespacing
\bibliographystyle{siam}
\bibliography{ref}
}

\appendix

\section{Carry-out Sequence}
\label{app-carryout}

The carry-out sequence $\mathbf{c}$
is generated
by the binary full-adder
operating over
an augend $\mathbf{a}$
and
an addend $\mathbf{b}$,
being
the resulting
sum the sequence
$\mathbf{s}$.
For example,
considering $n=3$,
we have the following
well-known structure:
\[
\begin{array}{cccccc}
&c_4 & c_3 & c_2 & c_1 & c_0
\\
&    & a_3 & a_2 & a_1 & a_0
\\
+&    & b_3 & b_2 & b_1 & b_0
\\
\cline{2-6}
&s_4 & s_3 & s_2 & s_1 & s_0
\end{array}
\]
The coefficients
of $\mathbf{c}$
are given by the following
recurrence:
\begin{align}
c_{i+1}
&
=
a_i \OpAnd b_i
\OpOr
c_i
\OpAnd
(a_i \BitXor b_i)
\\
&
=
a_i \OpAnd b_i
\OpOr
c_i
\OpAnd
(a_i \OpOr b_i)
,
\quad
i = 1,2,\ldots,n,
\end{align}
where $c_0 = 0$.
The sum coefficients
are given
by:
\begin{align}
s_i = c_i \BitXor a_i \BitXor b_i
,
\quad
i = 1,2,\ldots,n
.
\end{align}

The full adder
offers
an efficient way
of generating $\mathbf{c}$.
Its
truth table
is given
by:
\strut
\enskip
\begin{center}
\begin{tabular}{cccc}
\toprule
$c_i$ & $a_i$ & $b_i$ & $s_i$
\\
\midrule
0 & 0 & 0 & 0 \\
0 & 0 & 1 & 1 \\
0 & 1 & 0 & 1 \\
0 & 1 & 1 & 0 \\
1 & 0 & 0 & 1 \\
1 & 0 & 1 & 0 \\
1 & 1 & 0 & 0 \\
1 & 1 & 1 & 1 \\
\bottomrule
\end{tabular}
\end{center}
Therefore,
taking
$a_i$, $b_i$, and $s_i$
as input Boolean variables,
we have
that
\begin{align}
c_i
&
=
a_i \BitXor b_i \BitXor s_i
.
\end{align}
Algorithm~\ref{algo-getcarry}
describes the procedure.

\begin{algorithm}
\caption{Algorithm for solving the carry-out recursion}
\label{algo-getcarry}

\begin{algorithmic}[1]

\Procedure{getCarry}{$\mathbf{a}$, $\mathbf{b}$}

\State $\mathbf{s} \gets \mathbf{a} + \mathbf{b}$
\Comment{Addition efficiently computed}
\State $\mathbf{c} \gets \mathbf{a} \BitXor \mathbf{b} \BitXor \mathbf{s}$

\State \Return $\mathbf{c}$

\EndProcedure

\end{algorithmic}

\end{algorithm}

\end{document}